\documentclass{svjour3}                     
\smartqed  
\usepackage{graphicx}
\begin{document}

\title{Erratum to: An Entropy Functional for Riemann-Cartan Space-Times}


\author{F. Hammad}


\institute{F. Hammad \at
              D\'{e}partement ST, Universit\'{e} A. Mira, Route Targa Ouzemmour 06000, W. B\'{e}jaia, Algeria\\
              \email{fayhammad@gmail.com}}

\date{Received: date / Accepted: date}

\maketitle

\begin{abstract}
We correct the entropy functional constructed in \emph{Int. J. Theor. Phys. \textbf{51}, 362 (2012)}. The 'on-shell' functional one obtains from this correct functional possesses a holographic structure without imposing any constraint on the spin-angular momentum tensor of matter, in contrast to the conclusion made in the above paper.
\end{abstract}
\paragraph{}

The error made in \cite{1} was the missing torsion trace $Q_{i}={Q_{ij}}^{j}$ in the evaluation of the divergence terms. Indeed, the integral of a divergence in Riemann-Cartan space-times is of the form
$\int_{\mathcal{M}}\nabla_{i}V^{i}=\int_{\partial\mathcal{M}}n_{i}V^{i}-\int_{\mathcal{M}}2Q_{i}V^{i}$. Taking this into account, the entropy functional that should be considered is actually the following more 'economical' one
\begin{eqnarray}
\mathcal{S}=\int\mathrm{d}^{4}x\sqrt{-g}\Big[\alpha\big((\nabla_{i}u_{j})(\nabla^{j}u^{i})-(\nabla_{i}u^{i})^{2}\big)+(\lambda
g_{ij}+T_{ij})u^{i}u^{j}\qquad\qquad\nonumber
\\+\,(\Sigma_{ijk}+\Sigma_{ikj}+\Sigma_{kji})u^{i}\nabla^{j}u^{k}\Big].\label{eqn1}
\end{eqnarray}
The additional terms introduced in \cite{1} arise automatically when extracting total divergences. The variation of this functional with respect to the field $u^{i}$ reads
\begin{eqnarray}
\delta\mathcal{S}=\int\mathrm{d}^{4}x\sqrt{-g}\Big[2\alpha\big((\nabla_{i}\delta u_{j})(\nabla^{j}u^{i})-(\nabla_{i}\delta u^{i})(\nabla_{j}u^{j})\big)+2(\lambda g_{ij}+T_{ij})\delta u^{i}u^{j}\nonumber
\\+(\Sigma_{ijk}+\Sigma_{ikj}+\Sigma_{kji})\delta u^{i}\nabla^{j}u^{k}+(\Sigma_{ijk}+\Sigma_{ikj}+\Sigma_{kji})u^{i}\nabla^{j}\delta u^{k}\Big].\label{eqn2}
\end{eqnarray}
Extracting from the total divergences the boundary integrals which do not contribute to the variation, the condition $\delta\mathcal{S}=0$ reads
\begin{eqnarray}
\int\mathrm{d}^{4}x\sqrt{-g}\Big(2\alpha\nabla_{[i}\nabla_{j]}u^{j}+\,\big(2\alpha
Q_{i}g_{jk}-2\alpha Q_{k}g_{ij}+\Sigma_{ikj}\big)\nabla^{j}u^{k}\qquad\qquad\qquad\nonumber
\\+\,\big[\lambda g_{ij}+T_{ij}+\frac{1}{2}(\nabla^{k}+2Q^{k})(\Sigma_{ijk}+\Sigma_{kij}+\Sigma_{kji})\big]u^{j}\Big)\delta u^{i}=0.
\label{eqn3}
\end{eqnarray}
From the identity $2\nabla_{[i}\nabla_{j]}u^{j}=-R_{ij}u^{j}-2{Q_{ij}}^{k}\nabla_{k}u^{j}$, it follows that (\ref{eqn3}) is satisfied for arbitrary variations of $u^{i}$ if
\begin{eqnarray}
\big[2\alpha Q_{kij}+2\alpha Q_{i}g_{jk}-2\alpha Q_{k}g_{ij}+\Sigma_{ikj}\big]\nabla^{j}u^{k}\qquad\qquad\qquad\qquad\qquad\nonumber
\\-\,\big[\alpha R_{ij}-\lambda g_{ij}-T_{ij}-\frac{1}{2}(\nabla^{k}+2Q^{k})(\Sigma_{ijk}+\Sigma_{kij}+\Sigma_{kji})\big]u^{j}=0.\label{eqn4}
\end{eqnarray}
This in turn is satisfied for all $u^{i}$ if and only if the content of each of the two square brackets vanishes identically, leading straightforwardly as in \cite{1} to the Cartan-Sciama-Kibble field equations, but containing more correctly than in \cite{1} the metric $g_{ij}$ instead of the Kronecker-delta $\delta_{ij}$ in front of the torsion traces $Q_{i}g_{jk}$ and $Q_{k}g_{ij}$. The arguments used in \cite{1} to prove the uniqueness of the functional (\ref{eqn1}) remain unchanged. When the field equations are substituted in (\ref{eqn1}) after integration by parts, however, the 'on-shell' entropy functional becomes simply
\begin{eqnarray}
\mathcal{S}=\frac{1}{8\pi G}\int_{\mathcal{\partial M}}\mathrm{d}^{3}x\sqrt{|h|}n_{i}\big(u^{j}\nabla_{j}u^{i}-u^{i}\nabla_{j}u^{j}\big).
\label{eqn5}
\end{eqnarray}
So contrary to the conclusion made in \cite{1}, the holographic structure emerges without imposing any constraint on the spin-angular momentum tensor of matter. Using (\ref{eqn5}), the application and conclusions made in \cite{1} for the case of Dirac fields and black holes with intrinsic spin remain unchanged. For spin fluids obeying the Frenkel condition, however, the affine connection $\nabla$ in (\ref{eqn5}) reduces as in \cite{1} to the Levi-Civita connection $\mathring{\nabla}$ only for an isotropic deformation $u^{i}u^{j}=\frac{1}{4}u^{2}g^{ij}$.



\begin{thebibliography}{0}    
\bibitem{1} Hammad F.: Int. J. Theor. Phys. \textbf{51}, 362 (2012)
\end{thebibliography}
\end{document}


\title{An Entropy Functional for Riemann-Cartan Space-times}


\author{F. Hammad}


\institute{F. Hammad \at
              D\'{e}partement LMD-ST, Universit\'{e} A. Mira, Route Targa Ouzemmour B\'{e}jaia, W. B\'{e}jaia 06000, Algeria\\
              \email{fayhammad@gmail.com}}

\date{Received: date / Accepted: date}

\maketitle

\begin{abstract}
By viewing space-time as a continuum elastic medium and
introducing an entropy functional for its elastic deformations, T.
Padmanabhan has shown that general relativity emerges from varying
the functional and that the latter suggests holography for gravity
and yields the Bekenstein-Hawking entropy formula. In this paper
we extend this idea to Riemann-Cartan space-times by constructing
an entropy functional for the elastic deformations of space-times
with torsion. We show that varying this generalized entropy
functional permits to recover the full set of field equations of
the Cartan-Sciama-Kibble theory. Our generalized functional shows
that the contributions to the on-shell entropy of a bulk region in
Riemann-Cartan space-times come from the boundary as well as the
bulk and hence does not suggest that holography would also apply
for gravity with spin in space-times with torsion. It is
nevertheless shown that for the specific cases of Dirac fields and
spin fluids the system does become holographic. The entropy of a
black hole with spin is evaluated and found to be in agreement
with Bekenstein-Hawking formula.
\end{abstract}\keywords{Riemann-Cartan space-time \and Entropy functional \and Cartan-Sciama-Kibble field equations \and Black holes \and Holography.}

\section{Introduction}\label{intro}
The study of black hole thermodynamics is certainly one of the
most promising ways towards a deep understanding of the quantum
nature of space and time. The pioneering works of Bekenstein
\cite{1,2} and Hawking \cite{3,4}, based respectively on the study
of the properties of black holes geometry rising from general
relativity and a semiclassical combination of the latter with
quantum field theory, would later give birth to the novel concept
of space-time holography \cite{5,6}. It is then interesting to
investigate the role of any possible extension of general
relativity in the study of black hole thermodynamics in the hope
of learning more about the nature of space-time at the quantum
level.

One of the old extensions of general relativity is the
Einstein-Cartan theory of gravity. In this theory the intrinsic
spin of a particle is naturally included in the geometric
description of space-time using Riemann-Cartan geometry instead of
the Riemannian geometry, i.e. by introducing torsion besides the
metric as another degree of freedom for space-time. In this theory
the Einstein field equations are replaced by what are commonly
known as the Cartan-Sciama-Kibble (CSK) field equations, first
discovered by Cartan \cite{7}, then independently rediscovered
later by Kibble \cite{8} and Sciama \cite{9}.

Ironically, the initial idea that led \'{E}. Cartan to generalize
the Riemannian geometry and introduce the concept of torsion for
space-time came from an analogy with the macroscopic concept of
torsion in the physics of continuum media. This relation between
space-time and continuum media stayed a mere analogy, though, up
until 1967 when A. D. Sakharov proposed that general relativity
may after all be just a low-energy approximation to the dynamics
of space-time in the same sense that elasticity is an
approximation to the microstructure of solids \cite{10,11,12}.
Recently, various authors have investigated the idea of an
eventual elasticity of space-time either by applying concepts from
elasticity theory to explore its dynamics \cite{13,14,15} or by
bringing novel interpretations to some fundamental concepts of
modern cosmology such as inflation \cite{16,17,18} and cosmic
strings \cite{19,20,21} by generalizing the three dimensional
theory of defects to space-time.

Recently still, T. Padmanabhan has taken up this idea of elastic
space-time from a thermodynamic viewpoint and introduced an
entropy functional to be associated with the elastic deformations
\cite{22,23,24}. When extremized in accordance with the second law
of thermodynamics, the entropy functional yields the equations of
general relativity (see also \cite{25}). Furthermore, the
functional implies that the contributions to the on-shell entropy
of a bulk region of space-time reside only on the boundary of the
region in accordance with the holographic principle. In addition,
the entropy of spinless black holes is found to be proportional to
the area of their event horizon in exact agreement with
Bekenstein-Hawking formula. The ability of this approach to
reproduce both the fundamental equations of general relativity and
main results from the thermodynamics of black holes suggests that
extending the approach into Riemann-Cartan space-times may provide
a simple way to include spin in the study of black hole
thermodynamics. The question then is whether it would be possible
to construct a generalized functional that reproduces the full set
of the CSK field equations as well as familiar results from black
hole thermodynamics. It is our aim in this paper to show that this
is indeed possible.

The paper is organized as follows. In Sect. \ref{CSK} we introduce
an entropy functional for space-times with torsion and show how
the CSK field equations emerge from varying the functional with
respect to the deformation vector field. In Sect. \ref{OSEF} we
use the CSK field equations to obtain a general form for the
on-shell functional in which the boundary contributions are
separated from the bulk ones. We then examine two specific cases
of matter fields with spin embedded in space-time with torsion for
which the functional takes exactly the form obtained by
Padmanabhan for Riemannian space-times. Moving on to space-times
with an event horizon, we show in Sect. \ref{EBHS} how the
Bekenstein-Hawking entropy formula for a black hole with spin is
recovered in our approach. We conclude this work with a discussion
section to highlight and comment our main results.

\section{The CSK Field Equations from an Entropy Functional}\label{CSK}
The motivation behind the introduction of an entropy functional by
Padmanabhan \cite{22,23,24} was to consider, in the spirit of
Sakharov, space-time as a continuum medium subject at low-energy
to elastic deformations characterized by the differential
deformation vector field (or the displacement vector field)
$u^{i}(x)=\bar{x}^{i}-x^{i}$ ($i=0,...,3$), where $x^{i}$ and
$\bar{x}^{i}$ are coordinates in the elastic space-time before and
after deformation respectively. Matter, through its energy-moment
tensor $T_{ij}$, is viewed in this approach as a default inside
the medium that spoils translational invariance in the field
$u^{i}$. It contributes to the functional through a quadratic
coupling with $u^{i}$. Further, in order not to get more than
second-order differential equations when the functional is varied
with respect to $u^{i}$, the former is constrained to contain at
most first-order covariant derivatives of the latter. In \cite{25}
the same arguments were used to introduce a general expression for
the functional but a different philosophy was adopted to arrive at
its final form which we reproduce here to serve as the starting
point for our construction of a generalized version:
\begin{equation}
\mathcal{S}=\int\mathrm{d}^{4}x\sqrt{-g}\Big[\alpha\big((\nabla_{i}u_{j})(\nabla^{j}u^{i})-(\nabla_{i}u^{i})^{2}\big)+(\lambda
g_{ij}+T_{ij})u^{i}u^{j}\Big].\label{eqn1}
\end{equation}
The constant $\alpha$ was determined to be $1/8\pi G$ and the
scalar $\lambda$ was constrained, using one of the Bianchi
identities in Riemannian geometry, to be $\frac{1}{2}\alpha
R-\Lambda$ with $R$ being the Ricci scalar and $\Lambda$ a
cosmological constant.

In order to generalize the above functional for Riemann-Cartan
space-times in the presence of matter with non-zero spin we need
to take into account torsion, characterized by the tensor
${Q_{ij}}^{k}={\Gamma_{[ij]}}^{k}=\frac{1}{2}({\Gamma_{ij}}^{k}-{\Gamma_{ji}}^{k})$,
as an additional degree of freedom in the geometry of space-time
(${\Gamma_{ij}}^{k}$ being the affine connection.) On the other
hand, to handle matter with spin we know that, besides the
symmetric energy-momentum tensor $T_{ij}$, one must also take into
account the contributions of the spin-angular momentum tensor
$\Sigma_{ijk}$ which is antisymmetric in its first two indices.
Now, in our approach metric and torsion will be taken to be
emergent properties of space-time rather then being fundamental
dynamical quantities in the same sense proposed in \cite{22,23,24}
regarding the metric of Riemannian space-times. Therefore, instead
of the usual general relativistic definition of the spin-angular
momentum tensor $\Sigma_{ijk}$ based on the variation with respect
to torsion \cite{26,27,28}, it is the field-theoretic definition
\cite{29}
\begin{equation}
{\Sigma_{ij}}^{k}=\frac{1}{2}\frac{\partial\mathcal{L}}{\partial\partial_{k}\varphi^{a}}(M_{ij})^{a}_{\;\;b}\varphi^{b}\label{eqn2}
\end{equation}
that should be adopted in our approach. $(M_{ij})^{a}_{\;\;b}$
being $d_{r}\times d_{r}$ matrices in some representation $r$ of
the Lorentz group generators acting on the fields $\varphi^{a}$
\footnote{ Here we use a different positioning of indices on
$\Sigma_{ijk}$ from the one used in \cite{29}. Also, we shall use
through out this paper letters from the middle of the Latin
alphabet to denote holonomic space-time indices and letters from
the beginning of the alphabet to denote both anholonomic
tangent-space indices and group representation indices.}. From the
tensor $\Sigma_{ijk}$ one can build another tensor, also
antisymmetric in its first two indices, called the spin-energy
potential and related to $\Sigma_{ijk}$ by \cite{29}
\begin{equation}
\Psi_{ijk}=\Sigma_{ijk}+\Sigma_{ikj}+\Sigma_{kji}.\label{eqn3}
\end{equation}
Hence the spin property of matter gives rise to an additional
tensor that should also appear as, among other things, a
translational-invariance breaking term inside the functional, i.e.
in a term analogous to $T_{ij}u^{i}u^{j}$ in (\ref{eqn1}).

Now the way to construct a scalar functional quadratic in the
field $u^{i}$ and containing first-order derivatives of this
latter, using the various contributions coming from the tensors
mentioned above, is to judiciously contract the different indices
each tensor brings to the functional with those of the field
$u^{i}$ and its derivatives. To help us find the right combination
among the many possibilities, we start with the form displayed in
(\ref{eqn1}) and successively add terms brought by the new
ingredients in such a way that our generalized functional reduces
to (\ref{eqn1}) in the absence of spin.

Beginning with the translational-invariance breaking terms, the
analogue of the quadratic term $T_{ij}u^{i}u^{j}$ in (\ref{eqn1})
that could be constructed from $\Psi_{ijk}$, given its
antisymmetry in the first two indices, is $\Psi_{ijk}u^{j}u^{k}$
with still one more index to be contracted away. By dimensional
analysis, however, we know that $\partial_{i}{\Psi^{i}}_{jk}$ has
the dimension of an energy-momentum tensor \cite{29} while the
torsion tensor ${Q_{ij}}^{k}$ has the same dimension as that of a
derivative ($\mathrm{lenght}^{-1}$). So the natural analogue of
$T_{ij}u^{i}u^{j}$ would be $Q^{i}\Psi_{ijk}u^{j}u^{k}$, where
$Q_{i}={Q_{ij}}^{j}$ is the trace of torsion. Now there is still
the possibility of producing a quadratic term by contracting two
components of the field $u^{i}$ in an antisymmetric way with
$Q^{i}\Psi_{ijk}$. Indeed, if we introduce the oriented 2-surface
$w^{jk}$ constructed from the parallelogram formed by two
components $u^{j}$ and $u^{k}$, then $w^{jk}$ is quadratic in the
field $u^{i}$ and antisymmetric. Hence we shall add
$Q^{i}\Psi_{i[jk]}w^{jk}$ as a third translational-invariance
breaking term in the functional.

Next, note that in the absence of spin we had the quadratic term
$\lambda g_{ij}u^{i}u^{j}$ containing the contributions of the
scalar curvature and the cosmological constant through $\lambda$.
Since the additional feature brought into the geometry by torsion
is antisymmetry, we expect the analogue of $\lambda
g_{ij}u^{i}u^{j}$ to be a quadratic term that would rise as an
antisymmetric coupling with the field $u^{i}$ and whose origin
would be due to torsion. Hence we use again the oriented 2-surface
$w^{ij}$ and introduce the term $\theta_{ij}w^{ij}$. The tensor
$\theta_{ij}$ is antisymmetric, coming from torsion, and whose
explicit form should be determined using constraints from geometry
as we did for $\lambda g_{ij}$ in the absence of spin \cite{25}.
More specifically, we shall use again one of the Bianchi
identities but in Riemann-Cartan space-time.

Staying with the contributions of geometry, it is clear that other
ways of introducing torsion contracted with the field $u^{i}$
without the use of $\Psi_{ijk}$ still exist. To guide us in the
search for the right combination we simply proceed by building on
the 'kinetic' term in (\ref{eqn1}) consisting of the first two
terms of the integrand. Indeed, the simplest and most
straightforward way to introduce a coupling between the fields
${Q_{ij}}^{k}$ and $u^{i}$ is to perform, in a way reminiscent of
gauge theories, the following substitution
$\nabla_{i}\rightarrow\widetilde{\nabla}_{i}=\nabla_{i}-Q_{i}$
where $Q_{i}$ is the trace of ${Q_{ij}}^{k}$. Actually, a
satisfactory justification for this seemingly \emph{ad hoc}
substitution will shortly emerge below.

Finally, by dimensional analysis we also learn that contracting
the tensor $\Psi_{ijk}$ with one component $u^{i}$ and one
covariant derivative $\nabla^{j}u^{k}$ produces a scalar of the
same dimension as the term $T_{ij}u^{i}u^{j}$. Therefore we also
add into the functional the following covariant scalar
$\Psi_{ijk}u^{i}\nabla^{j}u^{k}$. In fact, when written in terms
of $\Sigma_{ijk}$ using (\ref{eqn3}), this term displays both a
symmetric and antisymmetric coupling with regard to the two
indices of $\nabla^{j}u^{k}$ as well as those of $u^{i}$ and
$u^{k}$. Furthermore, if we perform on this term the above
mentioned substitution
$\nabla_{i}\rightarrow\widetilde{\nabla}_{i}$ we see the emergence
of precisely the quadratic term $Q^{i}\Psi_{ijk}u^{j}u^{k}$ which
we introduced above from different motivations, so that all the
terms nicely hang together at the end.

Putting all these terms together into a generalized entropy
functional, this latter reads
\begin{eqnarray}
\mathcal{S}=\int\mathrm{d}^{4}x\sqrt{-g}\Big[\alpha\big((\widetilde{\nabla}_{i}u_{j})(\widetilde{\nabla}^{j}u^{i})-(\widetilde{\nabla}_{i}u^{i})^{2}\big)+(\lambda
g_{ij}+T_{ij})u^{i}u^{j}\qquad\qquad\nonumber
\\+\,(\Sigma_{ijk}+\Sigma_{ikj}+\Sigma_{kji})u^{i}\widetilde{\nabla}^{j}u^{k}+(\theta_{ij}-Q^{k}\Sigma_{ijk})w^{ij}\Big],\label{eqn4}
\end{eqnarray}
where we used the relation (\ref{eqn3}) to trade everywhere
$\Psi_{ijk}$ for $\Sigma_{ijk}$ in order to be able later to
compare our final results with those given in reference \cite{26}.
The detailed form we get when substituting the explicit expression
of $\widetilde{\nabla}$ is
\begin{eqnarray}
\mathcal{S}=\int\mathrm{d}^{4}x\sqrt{-g}\Big[\alpha\big((\nabla_{i}u_{j})(\nabla^{j}u^{i})-(\nabla_{i}u^{i})^{2}\big)+(\lambda
g_{ij}+T_{ij}+2Q^{k}\Sigma_{kij})u^{i}u^{j}\qquad\qquad\nonumber
\\+\,(\Sigma_{ijk}+\Sigma_{ikj}+\Sigma_{kji}+2\alpha Q_{i}\delta_{jk}-2\alpha Q_{k}\delta_{ij})u^{i}\nabla^{j}u^{k}+(\theta_{ij}-Q^{k}\Sigma_{ijk})w^{ij}\Big].\nonumber\\\label{eqn5}
\end{eqnarray}
In what follows we shall see that this functional reproduces both
groups of field equations of Cartan-Sciama-Kibble theory if we
demand that it be extremized for any deformation vector field
$u^{i}$. To perform correctly the variation of (\ref{eqn5}) with
respect to the field $u^{i}$, we symbolically write the oriented
2-surface $w^{ij}$ of the parallelogram as $u^{i}\wedge u^{j}$ in
order to take care of the oriented product. Thus we have $\delta
w^{ij}=\delta u^{i}\wedge u^{j}+u^{i}\wedge\delta
u^{j}=u^{i}\wedge\delta u^{j}-u^{j}\wedge\delta u^{i}$. When this
is done in the variation of the functional, the condition $\delta
\mathcal{S}=0$ for all $u^{i}$ yields
\begin{eqnarray}
2\alpha\nabla_{[i}\nabla_{j]}u^{j}+\,\big(2\alpha
Q_{i}\delta_{jk}-2\alpha
Q_{k}\delta_{ij}+\Sigma_{ikj}\big)\nabla^{j}u^{k}\qquad\qquad\qquad\nonumber
\\+\,\big[\lambda
g_{ij}+T_{ij}-\theta_{ij}-2\alpha\nabla_{[i}Q_{j]}+\frac{1}{2}(\nabla^{k}+2Q^{k})(\Sigma_{ijk}+\Sigma_{kij}+\Sigma_{kji})\big]u^{j}=0.
\label{eqn6}
\end{eqnarray}
Next, one may write the second-order derivatives of $u^{i}$ in
(\ref{eqn6}) as a combination of first-order and zero-order
derivatives if one uses the following contracted Ricci identity in
Riemann-Cartan geometry \cite{26}:
$2\nabla_{[i}\nabla_{j]}u^{j}=-R_{ij}u^{j}-2{Q_{ij}}^{k}\nabla_{k}u^{j}$.
Hence identity (\ref{eqn6}) becomes
\begin{eqnarray}
\big[2\alpha Q_{kij}+2\alpha Q_{i}\delta_{jk}-2\alpha
Q_{k}\delta_{ij}+\Sigma_{ikj}\big]\nabla^{j}u^{k}\qquad\qquad\qquad\qquad\qquad\nonumber
\\-\,\big[\alpha R_{ij}-\lambda
g_{ij}-T_{ij}+\theta_{ij}+2\alpha\nabla_{[i}Q_{j]}-\frac{1}{2}(\nabla^{k}+2Q^{k})(\Sigma_{ijk}+\Sigma_{kij}+\Sigma_{kji})\big]u^{j}=0,\nonumber\\\label{eqn7}
\end{eqnarray}
which, in turn, may be satisfied for any $u^{j}$ if and only if
each of the factors vanishes separately, that is
\begin{eqnarray}
2\alpha(Q_{ijk}+Q_{j}\delta_{ik}-Q_{i}\delta_{jk})=\Sigma_{ijk}\label{eqn8}
\end{eqnarray}
and
\begin{eqnarray}
\alpha R_{ij}-\lambda
g_{ij}=T_{ij}-\theta_{ij}-2\alpha\nabla_{[i}Q_{j]}+\frac{1}{2}(\nabla^{k}+2Q^{k})(\Sigma_{ijk}+\Sigma_{kij}+\Sigma_{kji}).\qquad\label{eqn9}
\end{eqnarray}
We recognize in (\ref{eqn8}) the first group of the CSK field
equations \cite{26}. Now the second group of field equations of
the CSK theory has also emerged and it resides in (\ref{eqn9}).
This can readily be seen as follows. From (\ref{eqn8}) one easily
extracts the following identities:\\
$\nabla^{k}\Sigma_{ijk}=4\alpha\nabla_{[i}Q_{j]}+2\alpha
\nabla^{k}Q_{ijk}$ and $Q^{k}\Sigma_{ijk}=2\alpha Q^{k}Q_{ijk}$.
Using these after taking the antisymmetric part of (\ref{eqn9})
and then comparing the result with the following Bianchi identity
in Riemann-Cartan geometry \cite{26}:
$R_{[ij]}=\nabla^{k}Q_{ijk}+2Q^{k}Q_{ijk}+2\nabla_{[i}Q_{j]}$,
immediately reveals what the antisymmetric tensor $\theta_{ij}$
is:
\begin{equation}
\theta_{ij}=-2\alpha\nabla_{[i}Q_{j]}.\label{eqn10}
\end{equation}
It is a tensor built from torsion as we argued above that it would
be. Finally, substituting (\ref{eqn10}) into (\ref{eqn9}), this
latter reads
\begin{eqnarray}
\alpha(R_{ij}-\frac{1}{2}g_{ij}R)+\Lambda
g_{ij}=T_{ij}+\frac{1}{2}(\nabla^{k}+2Q^{k})(\Sigma_{ijk}+\Sigma_{kij}+\Sigma_{kji}).\label{eqn11}
\end{eqnarray}
This is the second group of the CSK field equations comprising a
cosmological constant.

Having built an entropy functional that reproduces the full set of
the CSK field equations when extremized, our aim is to apply it to
the evaluation of the entropy of black holes with non-zero spin in
space-times with torsion. Before we do that, however, we shall
first derive in the next section a general form for the on-shell
functional and examine two specific cases of matter fields with
non-zero spin.

\section{The On-shell Entropy Functional}\label{OSEF}
The study of the entropy functional we are going to conduct in
this section will be on-shell, i.e. the CSK field equations will
be assumed to be satisfied in the bulk region of the space-time
considered, so that entropy is kept extremized in accordance with
the second law of thermodynamics. When the functional (\ref{eqn5})
is integrated by parts and both groups of the CSK field equations
(\ref{eqn8}) and (\ref{eqn11}) are taken into account along with
the constraint (\ref{eqn10}) on $\theta_{ij}$, the entropy
functional becomes
\begin{eqnarray}
\mathcal{S}=\frac{1}{8\pi G}\int_{\mathcal{M}}
\mathrm{d}^{4}x\sqrt{-g}\Big[\nabla_{i}\big(u^{j}\nabla_{j}u^{i}-u^{i}\nabla_{j}u^{j}-8\pi
G\Sigma^{i}_{\;jk}u^{j}u^{k}\big)\nonumber
\\-\,\big(2\nabla_{[i}Q_{j]}+8\pi
GQ^{k}\Sigma_{ijk}\big)w^{ij}\Big].\label{eqn12}
\end{eqnarray}
Using Stokes' theorem on the first parentheses and the Bianchi
identity on the second, the functional also takes the following
form
\begin{eqnarray}
\mathcal{S}=\frac{1}{8\pi G}\int_{\mathcal{\partial
M}}\mathrm{d}^{3}x\sqrt{|h|}n_{i}\big(u^{j}\nabla_{j}u^{i}-u^{i}\nabla_{j}u^{j}-8\pi
G\Sigma^{i}_{\;jk}u^{j}u^{k}\big)\nonumber\\
+\,\frac{1}{8\pi G}\int_{\mathcal{M}}
\mathrm{d}^{4}x\sqrt{-g}\big(\nabla^{k}Q_{ijk}-R_{[ij]}\big)w^{ij},
\label{eqn13}
\end{eqnarray}
where $h$ is the determinant of the three-dimensional metric
corresponding to the hypersurface $\partial\mathcal{M}$ bounding
the integration region $\mathcal{M}$ of space-time, and $n_{i}$ a
unit vector normal to that hypersurface. We therefore conclude
that the contributions to the on-shell entropy of a bulk region
$\mathcal{M}$ of Riemann-Cartan space-times do not reside
exclusively on the boundary $\partial\mathcal{M}$ but do partially
come from the bulk region $\mathcal{M}$ as well. We shall
nevertheless expose below two specific cases of matter fields for
which we recover an entropy whose contributions solely come from
the boundary $\partial\mathcal{M}$ and whose explicit expression
coincides with the one found by Padmanabhan for Riemannian
space-times. In what follows we shall first treat the case of the
spin-half Dirac field and then examine the case of spin fluids.

\subsection{The On-shell Functional in the Presence of a Dirac Field}\label{EFDF}
The Lagrangian of a free Dirac field $\psi$ in a curved space-time
with torsion is \cite{26}
\begin{equation}
\mathcal{L}=\frac{i}{2}(\overline{\psi}\gamma^{k}\psi_{;k}-\overline{\psi}_{;k}\gamma^{k}\psi)+\frac{i}{4}K_{jkl}\overline{\psi}\gamma^{[j}\gamma^{k}\gamma^{l]}\psi-m\overline{\psi}\psi,\label{eqn14}
\end{equation}
where
$\psi_{;k}=\partial_{k}\psi+\frac{1}{4}\omega_{kab}\gamma^{[a}\gamma^{b]}\psi$
with $\omega_{kab}$ being the Riemannian part of the tangent-space
connection and $K_{jkl}=Q_{jkl}+Q_{ljk}-Q_{klj}$ is the contortion
tensor. $\gamma_{i}$ are space-time Dirac gamma matrices related
to the constant tangent-space gamma matrices $\gamma_{a}$ through
the Vielbeins $\gamma_{i}=e_{i}^{\;a}\gamma_{a}$. Using the
definition (\ref{eqn2}) with
$M_{ij}=\frac{1}{2}\gamma_{[i}\gamma_{j]}$ for Dirac spinors
\cite{29}, one finds that the spin-angular momentum tensor is
given by
\begin{equation}
\Sigma_{ijk}=\frac{i}{4}\overline{\psi}\gamma_{[i}\gamma_{j}\gamma_{k]}\psi,
\label{eqn15}
\end{equation}
and hence is a totaly antisymmetric tensor. As a consequence, the
third term in the first parentheses of (\ref{eqn12}) vanishes. But
when using (\ref{eqn8}) to express torsion in terms of
$\Sigma_{ijk}$ it follows that $Q_{ijk}$ is also totaly
antisymmetric and hence traceless. So the content of the second
parentheses in (\ref{eqn12}) also vanishes. Finally, recalling
that in Riemann-Cartan space-times one uses in the covariant
derivative an affine connection $\Gamma_{ij}^{\;\;k}$ that is
related to the Christoffel connection $\{_{ij}^{\,k}\}$ by
\cite{26}
\begin{equation}
{\Gamma_{ij}}^{k}=\{_{ij}^{\,k}\}+{Q_{ij}}^{k}+Q^{k}_{\;\;ij}-{Q_{j}}^{k}\,\!_{i},\label{eqn16}
\end{equation}
one easily sees from the total antisymmetry of torsion that
$u^{j}\nabla_{j}u^{i}-u^{i}\nabla_{j}u^{j}=u^{j}\mathring{\nabla}_{j}u^{i}-u^{i}\mathring{\nabla}_{j}u^{j}$
where $\mathring{\nabla}$ is the covariant derivative in
Riemannian space-times. Hence, the form of the entropy functional
for a Riemann-Cartan space-time in the presence of a Dirac field
reduces to that of space-time without torsion \cite{22}
\begin{equation}
\mathcal{S}=\frac{1}{8\pi G}\int_{\mathcal{\partial
M}}\mathrm{d}^{3}x\sqrt{|h|}n_{i}\big(u^{j}\mathring{\nabla}_{j}u^{i}-u^{i}\mathring{\nabla}_{j}u^{j}\big).
\label{eqn17}
\end{equation}
The effects of torsion, and therefore of spin, are nevertheless
still present; being encoded inside the deformation vector field
$u^{i}$.

\subsection{The On-shell Functional in the Presence of a Spin Fluid}\label{EFSF}
Proceeding to the case of a space-time filled with a spin fluid we
shall use the following standard definition \cite{30,31}: By a
spin fluid we mean a perfect fluid of density $\rho$,
four-velocity $v^{i}$, and whose spin-angular momentum tensor
${\Sigma_{ij}}^{k}$ is given by
\begin{equation}
{\Sigma_{ij}}^{k}=\Xi_{ij}v^{k}. \label{eqn18}
\end{equation}
$\Xi_{ij}$ is the spin-density assumed to satisfy the so-called
Frenkel condition $\Xi_{ij}v^{j}=0$. Then from (\ref{eqn8}) we
have $Q_{i}=-2\pi G{\Sigma_{ij}}^{j}$, whence we immediately see
that the previous Frenkel condition produces a traceless torsion
$Q_{i}=0$. Integral (\ref{eqn12}) therefore reduces to
\begin{equation}
\mathcal{S}=\frac{1}{8\pi G}\int_{\mathcal{\partial
M}}\mathrm{d}^{3}x\sqrt{|h|}n_{i}\big(u^{j}\nabla_{j}u^{i}-u^{i}\nabla_{j}u^{j}-8\pi
G\Xi^{ij}v^{k}u_{j}u_{k}\big). \label{eqn19}
\end{equation}
Further, the explicit expression one obtains for $Q_{ijk}$ from
(\ref{eqn8}) and (\ref{eqn18}) is $4\pi G\Xi_{ij}v_{k}$. Then, on
using (\ref{eqn16}) we find that
$u^{j}\nabla_{j}u^{i}-u^{i}\nabla_{j}u^{j}=u^{j}\mathring{\nabla}_{j}u^{i}-u^{i}\mathring{\nabla}_{j}u^{j}+8\pi
G\Xi^{ij}v^{k}u_{j}u_{k}$ and so integral (\ref{eqn19}) gives back
(\ref{eqn17}) as the on-shell entropy functional for space-time
containing spin fluids.

\section{Entropy of Black Holes with Intrinsic Spin}\label{EBHS}
Having obtained the general form of the on-shell entropy
functional we now proceed to its evaluation in space-time with an
event horizon created by a black hole with non-zero spin, i.e.
either a rotating black hole or a static black hole with a
non-zero quantum spin. For that purpose, we shall adopt the idea
and follow the strategy that were introduced in \cite{22,23,24}
for treating event horizons in an approach that views space-time
as a continuum elastic medium. The idea is to view the forward
translation in time $t\rightarrow t+\varepsilon$ as a byproduct of
the deformation $u^{i}$, which reads $x^{i}\rightarrow
x^{i}+u^{i}$ on a spacelike hypersurface whose unit normal being
$u^{i}$, whereas the event horizon is viewed as a singular point
in the deformation field $u^{i}$ on which this latter satisfies
$n_{i}u^{i}=0$. The strategy then consists of evaluating the
functional in the Rindler-like space near the horizon.

According to (\ref{eqn8}), torsion is algebraically related to the
spin angular momentum so that torsion is vanishing outside the
sources. Hence integral (\ref{eqn12}) greatly simplifies when the
boundary enclosing the hole is taken to be the event horizon, for
then (\ref{eqn12}) reduces to
\begin{equation}
\mathcal{S}=\frac{1}{8\pi G}\int_{\mathcal{\partial
M}}\mathrm{d}^{3}x\sqrt{|h|}n_{i}\big(u^{j}\mathring{\nabla}_{j}u^{i})=\frac{1}{8\pi
G}\int_{\mathcal{\partial
M}}\mathrm{d}^{3}x\sqrt{|h|}n_{i}c^{i},\label{eqn20}
\end{equation}
where we have introduced the four-acceleration
$c^{i}=u^{j}\mathring{\nabla}_{j}u^{i}$ of the field $u^{i}$
\cite{22}. The next step is to go to the Rindler-like space near
the horizon to evaluate (\ref{eqn20}). For that, we need a
solution to the CSK field equations, which reduce to Einstein
vacuum equations outside the sources. Usually, when dealing with
vacuum Einstein equations one exploits the underling spherical
symmetry and adopts the Schwarzschild metric which is the unique
static spherical-symmetric solution according to Birkhoff's
theorem. Dealing with non-zero spin black holes, however, we don't
have spherical symmetry anymore but expect an axial-symmetric
solution whose axis would coincide with that of spin. So our
choice of space-time will be the Kerr space-time \footnote{This
choice of space-time is valid for a neutron black hole for
example. More generally though, it will be the Kerr-Newman
space-time when electric and/or magnetic charges are present. The
arguments we use and the results we obtain however are also valid
for these general cases.}. Note that, concerning black holes with
a quantum spin, our choice of space-time matches that of \cite{32}
where Kerr space-time is identified as a geometric model for the
spinning electron. Recall that the Kerr metric around a black hole
is specified by the mass $M$ and the angular momentum per unit
mass $a=J/M$ of the hole. For the case of a quantum spin, it is
$s$ that should play the role of the classical angular momentum
$J$, i.e. $a=s/M$. Now the Kerr metric in Boyer-Lindquist
coordinates may be written as follows \cite{33}
\begin{eqnarray}
&\mathrm{d}l^{2}&=-\frac{\varrho^{2}\Delta}{\xi^{2}}\mathrm{d}t^{2}+\frac{\varrho^{2}}{\Delta}\mathrm{d}r^{2}+\varrho^{2}\mathrm{d}\theta^{2}+\frac{\xi^{2}}{\varrho^{2}}\Big(d\phi-\Omega\mathrm{d}t\Big)^{2}sin^{2}\theta,\nonumber
\\\nonumber\\
&\varrho^{2}&=r^{2}+a^{2}cos^{2}\theta,\nonumber\\
&\Delta&=r^{2}-2GMr+a^{2},\nonumber\\
&\xi^{2}&=(r^{2}+a^{2})^{2}-a^{2}\Delta
sin^{2}\theta,\nonumber\\
&\Omega &=\frac{2aGMr}{\xi^{2}}.\label{eqn21}
\end{eqnarray}
From this metric which is singular for
$r_{\pm}=GM\pm\sqrt{G^{2}M^{2}-a^{2}}$ one deduces the radius of
the event horizon $r_{H}$, the area of the horizon $A_{H}$, and
the surface gravity $\kappa$
\begin{eqnarray}
&r_{H}&=r_{+},\nonumber\\
&A_{H}&=4\pi(r^{2}_{H}+a^{2}),\nonumber\\
&\kappa&=\frac{r_{+}-r_{-}}{4GMr_{+}}.\label{eqn22}
\end{eqnarray}
Before we proceed to the extraction of the Rindler-like metric, we
would like to make the following remark. It is well-known that
according to (\ref{eqn21}) one gets a naked singularity whenever
$a>GM$. So one might worry about such a singularity when dealing
with quantum spins given that for a Kerr electron \cite{32}, as
well as for a nucleus, one has $a=s/m\gg Gm$. For the case of
macroscopic black holes, however, we are dealing not with one
nucleus but with collapsed nuclear matter whose final spin
scarcely averages to a non-zero value $<s>$, so that the ratio
$s/M$ stays far below the dangerous limit of producing a naked
singularity.

Now, to extract the metric of the near-horizon region we follow
\cite{33} and expand (\ref{eqn21}) in terms of the variable
$N^{2}=4(r-r_{+})/(r_{+}-r_{-})$, keeping only the leading
contribution. Denoting by the subscript $H$ all quantities
evaluated on the horizon the Kerr metric takes, up to the
conformal transformation $g_{ij}\rightarrow\varrho^{2}_{H}g_{ij}$,
the following Rindler-like form \cite{33}
\begin{equation}
\mathrm{d}l^{2}\approx-N^{2}\kappa^{2}\mathrm{d}t^{2}+\mathrm{d}N^{2}+\mathrm{d}\theta^{2}+\frac{\xi_{H}^{2}}{\varrho_{H}^{4}}sin^{2}\theta
\mathrm{d}\tilde{\phi}^{2}=-\tilde{N}^{2}\mathrm{d}t^{2}+\frac{\mathrm{d}\tilde{N}^{2}}{\kappa^{2}}+\mathrm{d}\theta^{2}+\frac{\xi_{H}^{2}}{\varrho_{H}^{4}}sin^{2}\theta
\mathrm{d}\tilde{\phi}^{2},\label{eqn23}
\end{equation}
where $\tilde{N}=N\kappa$ and $\tilde{\phi}=\phi-\Omega_{H}t$.
Having this simple form of the metric we can now proceed,
following \cite{22}, to the evaluation of (\ref{eqn20}).
Remembering that the horizon sits at $N=0$, integral (\ref{eqn20})
reduces to
\begin{equation}
\mathcal{S}=\frac{1}{8\pi
G}\int_{0}^{\frac{2\pi}{\kappa}}\mathrm{d}\tau\int_{0}^{\pi}\int_{0}^{2\pi}
\tilde{N}(n_{i}c^{i})\xi_{H}sin\theta\mathrm{d}\theta
\mathrm{d}\tilde{\phi}, \label{eqn24}
\end{equation}
where we have integrated along the $\frac{2\pi}{\kappa}$-periodic
Euclideanized time $\tau=it$ that removes the conical singularity
\cite{33}. Next, we recall that for the Levi-Civit\`{a} covariant
derivative $\mathring{\nabla}$ one has on the horizon for which
$N\rightarrow 0$ the following limit
$\tilde{N}(n_{i}c^{i})\rightarrow \kappa$ \cite{22,23,24} (see
also \cite{34}.) Thereby, the above integral evaluates to
\begin{equation}
\mathcal{S}=\frac{\pi\xi_{H}}{G}=\frac{A_{H}}{4G}, \label{eqn25}
\end{equation}
thus recovering the Bekenstein-Hawking formula for black holes
with spin. In our approach, this last result is valid for black
holes with an intrinsic quantum spin as well as for black holes
with a classical spinning, i.e. the rotating black holes.

\section{Discussion}\label{D}
By extending the construction of the entropy functional for
space-times with torsion viewed as continuum elastic media, in
order to study their thermodynamics, we have built a functional
using arguments related to the physics of continuum media and
their defects. We saw that the functional remarkably reproduces
exactly both groups of the CSK field equations when demanding that
entropy be extremized in accordance with the second law of
thermodynamics for all possible deformations of space-time.
Furthermore, the new terms that appeared in the generalized
functional when torsion and spin are taken into account actually
add a nice intuitive interpretation for the rising of space-time
torsion from spin. Indeed, we see from (\ref{eqn5}) that in
addition to the quadratic coupling with the deformation vector
field through the simple product of two components of the latter,
the spin-angular momentum tensor has another quadratic coupling
that is orientation-dependent. More precisely, it couples to the
oriented 2-surface area formed by two components of the
deformation field. Thus, whereas the CSK field equations
(\ref{eqn8}) give a quantitative description of how space-time
torsion rises from spin, this new form of coupling in the
functional adds the following qualitative description. While
energy-momentum deforms space-time in analogy to a bowling ball
that bends a two-dimensional rubber sheet when placed on it,
spin-angular momentum deforms space-time by changing also the
orientation of each deformed infinitesimal area in analogy to an
elastic rod that torques when the relative orientation of its
edges gets twisted. The term $\Psi_{ijk}u^{i}\nabla^{j}u^{k}$
enhances this analogy since it represents a contribution that
evokes the "moment" of the deformation gradient.

Now, although every term of the functional (\ref{eqn5}) was
motivated using physical arguments, our construction remains
heuristic and the question of uniqueness of integral (\ref{eqn5})
has to be addressed here. In what follows we shall argue that
integral (\ref{eqn5}) is actually the unique functional that when
varied recovers exactly the CSK field equations (\ref{eqn8}) and
(\ref{eqn11}). In fact, suppose there are other terms that might
be added inside the integral to make it more general and still
leave unchanged equations (\ref{eqn8}) and (\ref{eqn11}). These
additional terms, if any, could only appear in the second and
third parenthesis of (\ref{eqn5}) since the structure of the first
parenthesis is dictated by the functional in the absence of spin
while the generality of the fourth parenthesis is already assured
by the tensor $\theta_{ij}$. Hence, the additional terms would be
of the form $A_{ij}u^{i}u^{j}$ and $B_{ijk}u^{i}\nabla^{j}u^{k}$,
where $A_{ij}(Q,\Sigma)$ would be a symmetric tensor built
quadratically (for dimensional reasons) from torsion and the
spin-angular momentum while $B_{ijk}(Q,\Sigma)$ would be a third
rank tensor linear in torsion and spin-angular momentum. When
these terms are added to (\ref{eqn5}) and this latter is varied,
one learns that in order for (\ref{eqn8}) and (\ref{eqn11}) to
remain valid, one should impose on $B_{ijk}$ and $A_{ij}$ the
following constraints: $B_{ijk}=B_{kji}$ and
$\frac{1}{2}\nabla^{j}B_{ijk}=A_{ik}$. Although the first of these
constraints is an algebraic one and may easily be achieved in the
framework of the CSK theory by a simple combination of $Q$'s and
$\Sigma$'s, the second constraint would bring an additional
dynamical equation to the theory, and hence could not be kept
without departing from the CSK theory in which dynamics emerges
solely from (\ref{eqn11}). Thus, we conclude that integral
(\ref{eqn5}) is the unique extension to Riemann-Cartan space-times
of the entropy functional capable in the framework of our approach
to lead exactly to the CSK theory.

This extension to Riemann-Cartan space-times of the functional
also brings a new insight on the holographic principle concerning
space-times with torsion. Indeed, in Sect. \ref{OSEF} we saw that
contrary to the case of zero-spin, the entropy of the bulk region
of space-time does not rise solely from the boundary of the region
but gets contributions from the bulk as well. Hence, the fact that
this same entropy when extremized yields the Einstein-Cartan
gravity theory without any prior assumption of the latter suggests
that, due to torsion, the holographic principle does not
necessarily apply for gravity with spin in Riemann-Cartan
space-times in general. We nevertheless saw two specific cases for
which the system is indeed holographic and displays the same
general formula for entropy found by Padmanabhan for Riemannian
space-times. Now, this unconformity with the holographic principle
is not due primarily to torsion created by the spin angular
momentum but more specifically to the trace of torsion that enters
the bulk integral in (\ref{eqn12}). Space-times with torsion are
thus holographic whenever the spin-angular momentum tensor of the
system is traceless. This observation helps us better understand
the relevance to holography of Dirac fields and spin fluids. Dirac
fields and spin fluids make space-times with torsion holographic
because of the tracelessness of their spin-angular momentum
tensors giving rise to a traceless torsion. For Dirac fields this
comes from the anticommutation of the gamma matrices whereas for
spin fluids it is thanks to the Frenkel condition
$\Xi_{ij}v^{j}=0$. Conversely, denoting by $\Sigma_{i}$ the trace
of the spin-angular momentum tensor $\Sigma_{ijk}$ then, since
$Q_{i}=-2\pi G\Sigma_{i}$, we arrive upon using (\ref{eqn12}) at
the following statement. If one ventures forth and adopts
holography as a first principle then one would conclude from this
that in space-times with torsion are allowed only systems whose
spin-angular momentum tensor is either traceless or constrained to
obey the following equation, $2\nabla_{[i}\Sigma_{j]}+8\pi
G\Sigma^{k}\Sigma_{ijk}=0$.

Finally, in Sect. \ref{EBHS}, applying our functional to
space-times with an event horizon we saw that the
Bekenstein-Hawking entropy formula is what is recovered for black
holes with a classical spinning and/or a quantum spin. In the
latter case, the explicit dependence on spin one gets from
(\ref{eqn25}) is
\begin{equation}
\mathcal{S}=2\pi\big(GM^{2}+\sqrt{G^{2}M^{4}-s^{2}}\big)\approx
4\pi GM^{2}-\frac{\pi s^{2}}{GM^{2}}+\mathcal{O}(s^{4}).
\label{eqn26}
\end{equation}
From this formula we see that in order for a black hole's quantum
spin to bring a notable difference with respect to the
thermodynamics of Schwarzschild black holes, the average quadratic
spin of the nucleons composing the black hole must be of order
$<s^{2}> \sim G^{2}M^{4}$. That is, if we denote by $n$ the number
of neutrons and by $n_{a}$ the number of the elementary spins
$\hbar$ aligned inside the neutron star before it collapses into a
black hole of mass, say, $3M_{\odot}$, the required fraction of
aligned spins is $\frac{n_{a}}{n}\sim\frac{3mGM_{\odot}}{\hbar
c}\sim10^{19}$ ($m$ being the neutron mass.) Thus, even if all the
nucleon spins inside the black hole where aligned, which is very
plausible in gravity with torsion given that spin-torsion
interaction induces spin alignment in the presence of strong
gravity inside collapsed matter \cite{35}, formula (\ref{eqn26})
shows that the quantum spin is far from being able to bring any
significant contribution to the thermodynamics of macroscopic
black holes. (The above required ratio $n_{a}/n$ exceeds unity
whenever the black hole's mass exceeds $\sim 10^{12}$kg.) Of
greater importance, however, is the fact that formula
(\ref{eqn26}) displays at first-order a departure from the entropy
of a Schwarzschild black hole by a term proportional to $<s^{2}>$.
This simple fact may actually be taken as a hint for the
consistency of our whole enterprize. Indeed, it is well-known
\cite{26,27,28,29} that the predictions of Einstein-Cartan gravity
theory depart from those of general relativity because of
quadratic terms coming from spin-spin contact interactions
$<\Sigma_{ijk}\Sigma^{ijk}>$. Therefore, formula (\ref{eqn26}) and
the reasoning that led to it represent, to some extent, an
indirect test that gives real credit to our inclusion of spin
\`{a} la Cartan in an elastic space-time.
